\documentclass[12pt]{article}
\usepackage{amssymb,amsmath,esint,titlesec}
\titleformat*{\section}{\normalsize\bf}
\titleformat*{\subsection}{\small\bf}

\begin{document}


\begin{titlepage}

\setlength{\baselineskip}{18pt}

                               \vspace*{0mm}

                             \begin{center}

{\Large\bf Time irreversibility from symplectic non-squeezing}

                            \vspace*{3.5cm}

              \large\sf  NIKOLAOS  \  \  KALOGEROPOULOS $^\dagger$\\

                            \vspace{0.2cm}
                            
 \normalsize\sf Center for Research and Applications \\
                                  of Nonlinear Systems \ (CRANS),    \\
   University of Patras, \  Patras 26500, \ Greece.                        \\

                            \end{center}

                            \vspace{3.5cm}

                     \centerline{\normalsize\bf Abstract}
                     
                           \vspace{3mm}
                     
\normalsize\rm\setlength{\baselineskip}{18pt} 

The issue of how time reversible microscopic dynamics gives rise to macroscopic irreversible processes 
has been a  recurrent issue in Physics since the time of Boltzmann whose ideas shaped, and essentially resolved,
such an apparent contradiction. Following Boltzmann's spirit and ideas, but employing Gibbs's approach, 
we advance the view  that  macroscopic irreversibility of Hamiltonian systems of many degrees of freedom 
can be also seen as a result of the  symplectic non-squeezing theorem.   \\

                           \vfill

                                                                         
                             \vfill

\noindent\rule{9cm}{0.2mm}\\  
   \noindent $^\dagger$ \small\rm Electronic address: \ \  \  {\normalsize\sf   nikos.physikos@gmail.com}\\

\end{titlepage}
 

                                                                                \newpage                 

\rm\normalsize
\setlength{\baselineskip}{18pt}

\section{Introduction}

There is little doubt that Gromov's symplectic non-squeezing theorem \cite{Gr1} is one of the most central results in Symplectic Geometry. 
This result is widely credited for concretely differentiating between symplectic and volume-preserving maps and, as a result, for 
establishing symplectic topology \cite{HofZ, McDS1} as an independent and free-standing line of research. 
The method of pseudo holomoprhic curves \cite{McDS2} that Gromov used to prove the non-squeezing theorem had a tremendous 
impact in several branches of Mathematics, as well as in String Theory, effects which are felt even today more 
than three decades after establishing that fundamental result. \\

Despite all this substantial impact in Mathematics, and String Theory, not many of its potential 
implications for Physics, and Statistical Physics in particular, have been explored, so far we know. 
A major exception that we are familiar with, are the works of M. de Gosson and his collaborators 
\cite{DeG1, DeG2, DeG3, DeG4, DeG5, DeG6}.      
His papers on the non-squeezing theorem have made accessible, to a typical Physics audience, the 
fundamental ideas contained in Gromov's and the subsequent  works on symplectic capacities \cite{HofZ}. \\

In this work, we present  a hand-waving argument purporting to show  that one can ascribe the macroscopic time irreversibility of  systems 
of many degrees of freedom,  whose microscopic dynamics is given by a Hamiltonian evolution, to the validity of the non-squeezing theorem.  
We largely follow Boltzmann's fundamental ideas on this issue, but use Gibbs's entropy, employing  
a few of the more recent results on the symplectic non-squeezing theorem and the symplectic capacities on this  matter.\\

In Section 2, we provide a few preliminaries about the symplectic non-squeezing theorem and symplectic capacities. 
Section 3 contains the main part of our argument.  Section 4 provides some conclusions and speculations. \\


\section{The non-squeezing theorem and symplectic capacities}

\subsection{Preliminaries}

Consider the $2n$-dimensional symplectic manifold \ $(\mathcal{M}, \omega)$. \ 
We recall that for \  $x\in\mathcal{M}$ \ and vectors \  $X,Y\in T_x\mathcal{M}$, \   \ $\omega$ \ is a non-degenerate 2-form
\begin{equation}
    \omega_x(X,Y) \ = \ 0, \ \ \ \forall \  Y \in T_x\mathcal{M} \ \ \implies \ \   X = 0    
\end{equation} 
which is also closed  
\begin{equation}   
     d\omega \ = \ 0 
\end{equation}  
Let \ $\mathfrak{L}_X$ \ stand for the Lie derivative and \ $i_X$ \ for the contraction along \ $X$. \ Then Cartan's formula gives
\begin{equation}  
     \mathfrak{L}_X \omega \ = \ d(i_X\omega) + i_X(d\omega)   
\end{equation}
which due to (2) reduces to 
\begin{equation}    
     \mathfrak{L}_X \omega \ = \ d(i_X\omega)
\end{equation}
Consider moreover $X_\mathcal{H}$ to be a Hamiltonian vector field: then, by definition \ $X_\mathcal{H} \in T\mathcal{M}$ \  is the 
generator of a Hamiltonian evolution/flow whose corresponding Hamiltonian function is \  $\mathcal{H}: \mathcal{M} \rightarrow \mathbb{R}$, 
where \ $X_\mathcal{H}$ \  is defined by  
\begin{equation}
   i_{X_{\mathcal{H}}} \omega \  = \ - d\mathcal{H}
\end{equation}
Substituting (5) into (4) we see that Hamiltonian vector fields preserve the symplectic form
\begin{equation}
   \mathfrak{L}_{X_\mathcal{H}} \omega = 0
\end{equation}
This relation can be considered as  a justification for the adoption of (2) in the definition of the symplectic form. 
The $2n$-form \ $\omega^n/n !$ \ is non-degenerate, so it can be used as the volume form of \ $(\mathcal{M}, \omega)$. \
Then
\begin{equation}   
      \mathfrak{L}_{X_\mathcal{H}} \left( \frac{\omega^n}{n !} \right) \  = \ 0
\end{equation} 
which is Liouville's theorem on the preservation of the symplectic volume under Hamiltonian flows. We see that the invariance of $\omega$ under 
Hamiltonian flows implies Liouville's theorem. Motivated by this realization, the question that arose was  whether there was actually any difference 
between symplectic and volume preserving diffeomorphisms of symplectic manifolds. The symplectic non-squeezing theorem \cite{Gr1} 
provided an affirmative answer. \\

A symplectic diffeomorphism \ $\phi: (\mathcal{M}, \omega ) \rightarrow (\mathcal{M}, \omega )$ \ is a diffeomorphism preserving the symplectic structure 
\begin{equation}
      \phi^\ast \omega \ = \  \omega
\end{equation}
Let \ $\omega_0$ \ indicate the standard symplectic form on \ $\mathbb{R}^{2n}$. \ Consider a local coordinate system 
 \ $(x_1,\ldots x_n, y_1,\ldots, y_n) \in \mathbb{R}^{2n}$ \  where, in the Hamiltonian Mechanics terminology, \ $y_i$ \  is the canonically 
 conjugate momentum to \ $x_i$ \ for \ $i=1,\ldots, n$. \  Then
\begin{equation}
     \omega_0 \ = \ \sum_{i=1}^n \ dx_i \wedge dy_i
\end{equation}
 By considering  \ $(\mathbb{R}^{2n}, \omega_0)$ \ instead of a general symplectic manifold $(\mathcal{M}, \omega)$,     one does not lose 
 any generality since, by Darboux's theorem, all  symplectic manifolds of dimension $2n$ are locally symplectically diffeomorphic 
 to \ $(\mathbb{R}^{2n}, \omega_0)$  \cite{HofZ, McDS1}.  \ Hence, and in stark 
contrast to the Riemannian case, all symplectic manifolds are locally ``equivalent": they can only be distinguished from each other by their 
dimension and by global invariants. \\


\subsection{The symplectic non-squeezing theorem}

In the terminology of the previous subsection, consider the open ball of radius \ $r$ \ in \ $\mathbb{R}^{2n}$ \ endowed with the Euclidean metric:  
\begin{equation}
    B_{2n}(r) \ = \ \{ \mathbb{R}^{2n} \ni (x_1, \ldots, x_n, y_1, \ldots, y_n): \ \ \ x_1^2 + \ldots + x_n^2 + y_1^2 + \ldots + y_n^2 \ \leq \ r^2 \}
\end{equation} 
Let \ $Z_{(x_1, y_1)} (R)$ \ be a cylinder of radius \ $R$ \ in \ $\mathbb{R}^{2n}$ \ based on the  symplectic 2-plane \ $(x_1, y_1)$: 
\begin{equation}
      Z_{(x_1, y_1)} (R) \ = \ \{ \mathbb{R}^{2n} \ni (x_1,\ldots x_n, y_1,\ldots, y_n):  \ \ \ x_1^2+y_1^2 \ \leq \ R^2 \} 
\end{equation}
The symplectic non-squeezing theorem \cite{Gr1} is the statement that \  $B_{2n}(r)$ \  can be embedded by a symplectic diffeomorphism 
in any cylinder based on a symplectic 2-plane, such as \ $Z_{(x_1,y_1)}(R)$, \ if \ $r \ \leq \  R$.  \ This is a non-obvious constraint and exists 
despite the fact that 
\begin{equation}  
     vol (B_{2n}(r)) \ < \ vol (Z_{(x_1, y_1)}(R))
\end{equation}
the latter volume being, obviously,  infinite. \\

Contrast this with the behavior of the volume-preserving maps: consider two compact domains 
\ $\Omega_1, \ \Omega_2 \subset \mathbb{R}^{2n}$ \ which are diffeomorphic, have smooth boundaries and equal volumes. 
Then \cite{DacM}  there is a volume-preserving diffeomorphism \ $\psi: \Omega_1 \rightarrow \Omega_2$. \    
Therefore, the additional obstruction provided by the symplectic non-squeezing theorem to symplectic embeddings, reveals the rigidity 
of the  symplectic embeddings, a property which is not shared by volume-preserving maps. Hence it is a sought-after difference between 
symplectic geometry and volume preserving/ergodic theory.    \\

A submanifold \ $(\mathcal{N}, \omega_{|\mathcal{N}})$ \ of a symplectic manifold \ $(\mathcal{M}, \omega)$ \ is called isotropic 
 if the restriction of the symplectic form \ $\omega$ \ on \ $\mathcal{N}$ \ is trivial, namely if \ $\omega_{|\mathcal{N}} = 0$. \  
 If the cylinder \ $Z$ \ is based on an isotropic 2-plane,
such as \ $(x_1,x_2)$ \ for instance, then no obstruction to symplectic embeddings exists; the ball \ $B_{2n}(r)$ \ can always be 
embedded in \ $Z_{(x_1,x_2)}(R)$ \  regardless of the relation between \ $r$ \ and \ $R$. \
Since a Hamiltonian vector field preserves the symplectic structure (6), it generates a one-parameter family of symplectic diffeomorphisms 
of \ $\mathbb{R}^{2n}$ \ which are trivially symplectic embeddings. Therefore Hamiltonian vector fields obey the symplectic 
non-squeezing theorem. \\

A coordinate-independent reformulation of the non-squeezing theorem  \cite{EG} states that a two-dimensional projection (``shadow")  of \ $B_{2n}(r)$ \ 
on a symplectic 2-plane has an area which is at least \ $\pi r^2$. \ To be more precise, consider a symplectic embedding 
\ $\varphi: B_{2n}(r) \rightarrow \mathbb{R}^{2n}$. \ Let \ $P$ \  indicate the orthogonal projection operator
onto the symplectic 2-plane \  $(x_1, y_1)$ \ and let \ $A$ \ stand for the area of a 2-dimensional measurable set on the 2-plane $(x_1, y_1)$.
Then the symplectic non-squeezing theorem states that
\begin{equation}  
    A (P\varphi (B_{2n}(r))) \ \geq \ \pi r^2
\end{equation}
A similar formulation exists by using the symplectic orthogonal, instead of the metric orthogonal projection on the symplectic 2-plane \ $(x_1,y_1)$. \\


\subsection{Symplectic capacities}

The essential features of symplectic maps distinguishing them from volume-preserving maps are distilled and axiomatized 
in the concept of ``symplectic capacities".  
Let \ $\mathfrak{M}_{2n}$ \ indicate the set of $2n$-dimensional symplectic manifolds. Then a symplectic capacity \ $c$ \ is a map \ 
$c: \mathfrak{M}_{2n} \rightarrow [0, +\infty]$ \ obeying the following three conditions:
\begin{itemize} 
    \item[--] {\sf Monotonicity:} For a symplectic embedding \ $f: (\mathcal{M}_1, \omega_1) \rightarrow (\mathcal{M}_2, \omega_2)$ \ 
                                                    between two symplectic manifolds, a symplectic capacity obeys
                                                            \begin{equation}
                                                                  c(\mathcal{M}_1, \omega_1) \ \leq \ c(\mathcal{M}_2, \omega_2) 
                                                            \end{equation} 

    \item[--] {\sf Conformality:} For a symplectic manifold \ $(\mathcal{M}, \omega)$ \ and for \ $\lambda\in\mathbb{R}\backslash \{ 0 \}$
                                                              \begin{equation}
                                                                  c(\mathcal{M}, \lambda\omega) \ = \ \lambda^2  \ c(\mathcal{M}, \omega)
                                                              \end{equation}
    \item[--] {\sf Normalization:} Following the notation of the non-squeezing theorem in \ $(\mathbb{R}^{2n}, \omega_0)$:
                                                             \begin{equation}
                                                                  c(B_{2n}(r=1)) \  = \ c(Z_{(x_i, y_i)} (R=1)) \ = \ \pi
                                                             \end{equation}
\end{itemize}
It turns out that the symplectic capacities are invariant under symplectic diffeomorphisms and  they are different from the volume due 
to the normalization condition (16). They measure the symplectic ``size" of a symplectic manifold. 
If one defines the Gromov width \ $c_G$ \ of the symplectic manifold  \ $(\mathcal{M}, \omega)$ \ as
\begin{equation}
    c_G (\mathcal{M}, \omega) \ = \ \sup_{r} \{\pi r^2 : \  \exists \ (B_{2n}(r), \omega_0) \stackrel{s}{\hookrightarrow} (\mathcal{M}, \omega)  \}
\end{equation}
where \ $\stackrel{s}{\hookrightarrow}$ \ stands for ``symplectic embedding", then  due to the non-squeezing theorem, \ $c_G$ \ is the smallest 
of all symplectic capacities, namely 
\begin{equation}
       c_G (\mathcal{M}, \omega) \ \leq \ c (\mathcal{M}, \omega)
\end{equation}
for all \  $(\mathcal{M}, \omega) \in \mathfrak{M}_{2n}$ \ and for all symplectic capacities \ $c: \mathfrak{M}_{2n} \rightarrow [0, +\infty ]$. 
Even though the explicit construction of symplectic capacities has proved to be non-trivial task, several such capacities have been constructed, so far. 
For a comprehensive, but non-exhaustive, list see \cite{CHLS}. \\   

Despite the above progress in symplectic geometry, very little of its body of knowledge has become known or used in the Statistical Physics literature.  
Classical Statistical Physics deals with systems having many  degrees of freedom. Explicit computation of symplectic capacities has proved to be 
a difficult task even for the simplest of spaces such as 4-dimensional ellipsoids \cite{Schlenk}. The case of 2-dimensional symplectic 
manifolds is in some sense not very interesting, and quite well-understood \cite{Sib, Jiang}. 
Given this, determination of symplectic capacities in high-dimensional symplectic manifolds which are of interest to Statistical Physics seems to 
be a hopeless task.  However the large number of degrees of freedom also brings along a significant simplification: in particular, the law of large 
numbers can be seen to imply that a locally Lipschitz function is almost constant in \ $\mathbb{R}^{2n}$ \ for \ $n\rightarrow\infty$. \ 
This is a substantial simplification when compared to the generic case of functions on \ $\mathbb{R}^{2n}$. \   
This idea lies at the core of the stability of results and of the predictive power  of Statistical Physics \cite{Gal, Gr-book}. \\

At our current level of understanding, it is not feasible to  make generic  statements  about symplectic capacities on arbitrary symplectic 
manifolds of high dimension that may be of physical interest. So one has to settle with less: there is a conjecture that all symplectic capacities 
are equivalent in high dimensional convex bodies (convex sets with non-empty interiors) in \ $\mathbb{R}^{2n}$. \ This is a conjecture, which if true, 
may have far-reaching consequences for both convex geometry, functional analysis and symplectic geometry. For some recent progress, 
from a particular viewpoint, one may consult \cite{AAKO, GO1}. This conjecture, if true, may also have physical consequences that are hard
 to foresee at this time.\\


\section{Time irreversibility and symplectic embeddings}

\subsection{Boltzmann and Gibbs entropies}

The issue of how a time-reversible microscopic dynamics (Hamilton's equations) can give rise to macroscopic irreversibility occupied a good part of the Physics research 
activity during the waning years of the 19th century. The basic ideas that provide a resolution to this apparent paradox were put forth by W. Thomson, J.C. Maxwell and L. 
Boltzmann. It is notable that even in the early 21st century, not all practitioners accept their explanations and some follow alternatives such as that, most 
notably, proposed by I. Prigogine and its school. 
We find the viewpoints and explanations contained in \cite{Leb1, Leb2, MN} who follow the Thomson/Maxwelll/Boltzmann views to be quite convincing on this 
particular issue.  As it befits such a fundamental issue in Physics, it is not surprising that there are still discussions about its logical and philosophical 
foundations \cite{Goldstein, Lavis}. Moreover, the many technical issues that have arisen in rigorously supporting this explanation have only be partly resolved \cite{Lan}.\\

The origin of the macroscopic time irreversibility is an issue where one has to carefully distinguish between the Boltzmann and the Gibbs views on entropy 
\cite{Jaynes, Leb1, Leb2, MN, Goldstein, Lavis}.  Boltzmann's expression 
\begin{equation}
      \mathcal{S}_B (\Gamma) \ = \ k_B  \log \Omega (\Gamma) 
\end{equation}
where \ $k_B$ \ stands for the Boltzmann constant and \ $\Omega (\Gamma)$ \ is the number of microscopic states corresponding to the macroscopic state \ $\Gamma$. \
This definition can easily accommodate the variation of entropy with time as well as describe the entropy of systems out of equilibrium \cite{Leb1, Leb2, MN}.  
Its main drawback is that it is very hard to perform explicit computations of \ $\mathcal{S}_B$. \ For explicit computations, people usually employ  the Gibbs entropy 
\begin{equation}
      \mathcal{S}_G [\rho] \ = \ - k_B  \int_\Omega \ \rho(x) \log \rho(x) \ dvol_\Omega
\end{equation}  
which reduces to \ $\mathcal{S}_B$ \ for a constant \ $\rho$ \ on $\Omega$. \ Here, 
 \ $\rho$ \ is a (Radon-Nikod\'{y}m) probability density in phase space \ $\Omega$, \ whose volume element is indicated by \ $dvol_\Omega$. \
Even though \ $\mathcal{S}_B$ \ and \ $\mathcal{S}_G$ \ have very similar functional forms as can be seen in their definitions (19), (20), they are quite different objects
expressing quite different viewpoints on exactly what entropy is \cite{Leb1, Leb2, MN, Goldstein, Lavis}. The Gibbs entropy (20), for instance, cannot change with time
under the Hamiltonian evolution of a system. This is a direct conclusion of Liouville's theorem (7). To allow (20) to change with time, one has to invoke some 
mechanism that plays an additional external role in the evolution of the system, such as noise due to thermal or quantum fluctuations, a coupling to an external reservoir etc. \\

The most widespread explanation for the increase of (20) with time however is ``coarse-graining". 
This amounts to the declaration of many features of the underlying system as not 
being known with absolute precision or to assuming such a precision as largely irrelevant so far as the collective properties of this system are concerned. The way to 
reconcile (20) with entropy variations (entropy increase) is to assume that the volume of phase space, which remains invariant under the Hamiltonian flow, undergoes a  
dramatically change in its shape under such a flow. 
The distortions of its shape are ultimately so severe that the distorted phase space volume becomes essentially indistinguishable from the whole 
volume of the phase space.  The ambient space and the distorted volumes  are indistinguishable, if someone looks at them with arbritrarily small but finite precision
\cite{Gal, Gr-book}. As a result the effective volume that he system occupies in phase space appears to increase, hence its entropy increases too.  
How this coarse-graining can occur as well as its logical, philosophical and technical foundations have been a topic of much discussion over the decades 
since tis inception \cite{V-book}. We explored  this viewpoint and the relation of coarse-graining of the phase space to entropy, mostly in a linear 
and convex geometric setting, in \cite{NK1}.\\ 

 A  motivation for considering coarse-graining for the increase of the Gibbs entropy with time, can be partially attributed to the Poincar\'{e} Recurrence Theorem \cite{KH, Zehn}.  
 The non-rigorous approach which invokes coarse-graining to justify the increase of  the entropy \ $\mathcal{S}_G$ \ with time has not really been rigorously
 proved in any  but he simplest of toy models \cite{Lan}. Such a derivation is based on the Boltzmann-Grad equation. 
 Its further application to the derivation of kinetic equations, especially to the case of  plasmas where the Vlasov equation \cite{Vlasov} is employed, 
 has been a topic of intense research activity recently. The approach of deriving these hydrodynamic equations will be encountered again in the sequel. \\

Contrary to  what was known until the advent of the non-squeezing theorem (ca. 1985), 
this result provides an obstruction to  arbitrary shape distortions of the phase 
space volume which are needed to justify the increase of \ $\mathcal{S}_G$. \     
The non-squeezing  theorem provides an additional constraint/rigidity of the phase space volume on top of Liouville's theorem (7), which questions the ``oil in water" 
spreading of this volume that has been evoked in this approach to justifying the increase of $\mathcal{S}_G$ with time.   
What exactly might the implications be of such a rigidity for Statistical Mechanics has been unclear up to this moment. 
We point out in this work one of the possible consequences of such a rigidity.\\

There are two diametrically opposite ways to argue what might the impact be of the symplectic non-squeezing theorem for Statistical
Mechanics: the first is to state that such a constraint is important enough, so that the phase space shape rigidity that it introduces significantly 
affects some of the results, even though such a behavior may not be easy 
to detect  or may be irrelevant for simple enough systems that we usually analyze.  However in more ``complex systems" where additivity becomes a non-trivial  
issue and which may be described  by power law or  entropies of other functional forms that are different from the Boltzmann/Gibbs/Shannon one, such as 
the Tsallis/q-entropy \cite{Ts-book}, or the $\kappa$-entropy \cite{Kan}, for instance,  such a rigidity may become quite important for their thermodynamic behavior. \\  

 The opposite view is that, maybe due to the large number of degrees of freedom which results in a substantial freedom of deformations, 
 or to the small rigidity of the phase space volume deformations introduced by the symplectic non-squeezing theorem, 
 its effects on the collective behavior of the system under study may be negligible.  As a result, the constraints that it imposes can be safely ignored in the 
 thermodynamic limit.  After all, this has been the case so far, long before the non-squeezing theorem was known; the coarse-grained picture
based on arbitrary phase  space volume deformations seemed to work quite well, even if not totally rigorously justified, 
without having any idea about the existence of the symplectic non-squeezing theorem. \\
             
      
 \subsection{The ``shape" of initial conditions and symplectic embeddings}      

In Gibbs's view of Statistical Mechanics, a system evolves under a given Hamiltonian, but to extract the  
ensemble of interest, the initial conditions have to be allowed to slightly deviate from the ones of any specific system. This 
does not create any problems in analyzing the behavior of the system under study, as dynamical systems with strong mixing 
properties lose track of  their initial states very fast \cite{Gal, KH}. Moreover, by considering systems having slightly 
different initial conditions from the given system and averaging over such initial conditions, one addresses the nagging 
problem about non-typical initial conditions that may  lead to highly improbable and, therefore, effectively non-observable outcomes. 
In such a case the averaging process renders such a-typical initial conditions, which have measure zero,  irrelevant
to the macroscopic behavior of the system \cite{Gal, Leb1, Leb2}. \\

So, it may be worth exploring what sets of initial conditions may be ``reasonable" for someone to consider. 
Various answers can be provided to this question depending on one's goals. One answer
pertains to the origin and the form of the  initial conditions of the Universe. This is a fundamental problem in Physics,
possibly to be resolved by Quantum Gravity,  and far outside the scope of the present formalistic work \cite{Leb1, Leb2}. 
We are far more modest: we aim  toward understanding physically relevant ``generic" initial 
conditions, that are also amenable to analytic calculations, or can be part of an analytic argument. 
We steer clear of the fundamental questions posed by Quantum Gravity, in this work, having in mind typical toy, 
point particle, models usually encountered in Statistical Physics.
Our approach involves making, sometimes severe, compromises between what is desirable and what is feasible. 
In our case some of the arguments will either apply to linear (as opposed to fully nonlinear) symplectic maps or to 
particular classes such as convex (as opposed to any shape) sets of initial conditions.  \\ 

Analytic computations pertinent to models of many degrees of freedom, excluding some more recent ideas employing 
entropic functionals that are not of the Boltzmann/Gibbs/Shannon form, are almost always  perturbations of   
Gaussians. This can be easily seen in the computations performed using the canonical or grand canonical ensembles 
which are, arguably, the most widely used approaches when one performs explicit calculations about the macroscopic properties 
of such systems \cite{Gal}. The corresponding Hamiltonians are perturbations of the classical harmonic oscillator which
is the prime example of a periodic system. The pertinent phase space curves expressing such a periodic motion 
are ellipsoids, which are essentially rescaled balls along some of their axes.\\

At the other end of the spectrum of the relevant physical behaviors, chaotic systems involve exponential sensitivity 
to initial conditions giving rise to rather complicated phase portraits. The initial conditions that can be relevant
in this case are slight perturbations of a particular orbit which expresses the Hamiltonian evolution of the system under 
study. The easiest way to proceed with such initial conditions is to consider them as ``equally spread", being points of a 
small disk in configuration space, which is perpendicular to the evolution of the system. The evolution of such a disc is a tube, 
at least for small values of the evolution parameter (``time").  
This is the strategy followed  in a Lagrangian approach to Mechanics employing a metric, rather than in the symplectic
 approach used in Hamiltonian Mechanics, see for  instance \cite{Pettini}.\\
 
What emerges from the above considerations is that there are at least two sets of  phase space ``shapes"   
that are physically pertinent and analytically manageable: polydiscs and ellipsoids.  
Let \ $D(a)$ \ denote an open disc centered at the origin and of radius \ $a>0$ \ in \ $\mathbb{R}^2$ \  endowed 
with the Euclidean metric. Parametrize \ $\mathbb{R}^{2n}$ \  as in (10), (11) above. 
Then the open polydisc \ $P(a_1, \ldots, a_n)$ in \  $\mathbb{R}^{2n}$ \ whose projections on 
the symplectic 2-planes \ $(x_i, y_i), \ i=1,\ldots n$ \  are \ $D(a_i), \ i=1,\ldots, n$ \  is      
\begin{equation}   
       P(a_1, \ldots, a_n) \ = \ D(a_1) \times \ldots \times D(a_n) 
\end{equation}
The polydisc encodes the shape of  the phase curves of a system of $n$-decoupled harmonic oscillators whose 
phase space is \ $\mathbb{R}^{2n}$. \ If \ $a_1 = \ldots = a_n = a$, \ then the polydisc reduces to the cube
\begin{equation}     
        Q(a) \ = \ P(a, \ldots, a) 
\end{equation}
Polydiscs therefore encode the simplest of the initial conditions as they rely on decoupled harmonic oscillators:
the individual degrees of freedom do not have time to interact with each other, let alone thermalize. On the the other hand 
one has the ball \ $B_{2n}(a)$ \ or, more generally, the ellipsoid 
\begin{equation}
       E(a_1, \ldots a_n) = \left\{(x_1,\ldots x_n, y_1, \ldots y_n): \ \  \sum_{i=1} ^n \  \frac{\pi (x_i^2 +y_i^2)}{a_i} \ < \ 1  \right\}
\end{equation}
which expresses the evolution of harmonic degrees of freedom that are coupled to each other with the only constraint being 
the their total energy to be less than one properly normalized unit.  
The ellipsoid expresses a fully thermalized system of harmonic oscillators 
having frequencies equal to the ellipsoid's lengths of semi-major axes along each of the $n$ complex directions. 
Notice that the projection of \  $E(a_1,\ldots, a_n)$ \  to the 2-plane \ $(x_i, y_i), \ i=1, \ldots, n$ \ is, obviously, \ $D(a_i), \ i=1,\dots, n$.\\

In some, very restricted, sense the above phase space ``shapes" that can be used to encode the evolution of pertinent 
initial conditions of Hamiltonian evolutions are as far away from each other as possible.  Indeed, consider initial conditions
whose ``shapes" are not disks but convex polyhedra, in general. Then, choosing a reasonable metric in the space of such initial
conditions such as the Banach-Mazur metric, it turns out that these conditions are as far from each other as possible. 
Within the limitations of a linear approach and within the restricted context of only allowing for convex combinations of 
initial conditions, the cube and the ball are as far from each other as possible. This convexity viewpoint and its implications 
for the phase space coarse-graining which conjecturally gives rise to  entropy was advanced in \cite{NK1}.\\     

The central point of the argument goes as follows: according to the symplectic non-squeezing theorem, the area of the projection 
(``shadow") of a symplectic ball on a symplectic 2-plane does not decrease under a symplectic map (13). Hamiltonian flows are 
symplectic maps as (6) shows, therefore (13) holds for Hamiltonian flows. Let us consider only the class of  Hamiltonian flows 
for which the area increases. This class of Hamiltonian flows is difficult to characterize from first principles. As will be seen in  subsection 3.4 in the 
sequel, for a system of many degrees of freedom, it does not appear unreasonable to expect that such an increase in the area of the projection on  
of the symplectic volume on a 2-dimensional symplectic plane will take place, above and beyond its change of shape during the flow. 
Such a change of shape during the flow will bring us outside the scope of validity of our argument, as the shape of the initial conditions will 
eventually cease to be a ball or even a convex body. However it may not be unreasonable to expect that for small times and upon
statistical averaging the overall shape of initial conditions will remain almost spherical/ellipsoidal, something that allows us to proceed.\\    
 
 Under such a projection on a 2-dimensional space 
where the 2-dimensional area  increases for Hamiltonian flows, the forward and backward directions of time are distinct: if the projection areas do not remain
invariant under temporal evolutions, they have to increase in the forward time direction, in the future. By time reversal, this means that the area  will have to 
decrease for the backward time direction, for the past. As a result the time reversed direction is not symplectic, as it 
violates the non-squeezing theorem, therefore it cannot be Hamiltonian. 
Hence any Hamiltonian flow even at a purely classical level can distinguish between the future and the past, as long as      
the areas of the 2-dimensional projections of the phase space volumes do not remain invariant under it.\\

One could wonder at this point what may the physical meaning of  the increase of the areas of the 2-dimensional projections on symplectic 2-planes
of the phase space volume.  Such projections cannot have any purely thermodynamic meaning, as  these 2-dimensional symplectic planes 
is parametrized by the microscopic rather than by the thermodynamic variables. The answer becomes obvious when one looks at the 
probability density function employed in Gibbs's approach and its marginals. 
The probability distribution function in the Gibbsian approach, is a function of the phase space coordinates \ 
$\rho_{2n} \ =  \ \rho \ (x_1, \ldots, x_n, y_1, \ldots, y_n)$. \ Determining such a function is a non-trivial task in general. To accomplish this one 
has to either evoke additional assumptions such as the ergodic hypothesis and/or to resort to several sets of approximations. \\

 One approach is to consider instead of the full probability distribution, its marginals which depend on the coordinates of few particles.   
 Then the effect of all the other particles
is averaged out giving rise to a ``background" that is assumed to vary much slower than the degrees of freedom under investigation.   
This assumed separation of scales between the effective background and the individual degrees of freedom of interest has proved 
to be a physically valid assumption, and is one of the main reasons why approaches that depend on 
such reductions are so effective. One could mention as examples of this approach, in various degrees,  the Boltzmann equation, the BBGKY
hierarchy and kinetic equations, such as the Vlasov equation, which use the marginals of \ $\rho_{2n}$, \   
most frequently the one-particle reduced probability function \  $\rho (x_1, y_1)$ \cite{HMcD}. \\\

 The one particle probability density function is the marginal
\begin{equation}   
         \rho_2 (x_1, y_1) \ = \ \int_{\mathbb{R}^{2n-2}}  \rho_{2n} (x_1,x_2,\ldots, x_n,y_1,y_2,\ldots,y_n) \ dx_2 \ldots dx_n dy_2\ldots dy_n 
\end{equation}
Probability marginals are the analytic analogues of projections in geometry. Hence the reduction of the full probability distribution function \  
$\rho_{2n}$ \  in phase space to its one-particle marginal, is essentially the projection of the $2n$-dimensional phase space volume 
to a $2$-dimensional symplectic plane.  The analogy becomes exact in convex geometry, and is extensively used 
in the Br\"{u}nn-Minkowski theory in particular \cite{Schn}. To reach a pertinent result in thermodynamics, one has to average the physically relevant
microscopic  quantities over all such projections. This can only be attained in relatively simple cases having  geometric and  potential physical interest.   
For recent results in asymptotic convex geometry (for very high $n$), which may potentially used to encode physically interesting thermodynamic behavior, 
one may consult  \cite{GO2}.\\


\subsection{Sections, projections and  intermediate symplectic capacities}

According to the symplectic non-squeezing theorem, the Hamiltonian flows either keep the areas of the symplectic 2-dimensional projections of spheres 
invariant,  or these areas have to increase with such flows. And this distinguishes the forward  from the backward flow (``time") direction. This argument 
would not work had someone used 2-dimensional sections of the volume of phase space instead of its symplectic 2-dimensional projections. 
The reason is well-understood \cite{AbbMatv}, but only in the linear setting: there is no such lower bound for sections under linear symplectic maps. 
To  be more precise  \cite{AbbMatv}, let \ $\phi: \mathbb{R}^{2n} \rightarrow \mathbb{R}^{2n}$ \ be a linear symplectic diffeomorphism, and let \ 
$V\subset\mathbb{R}^{2n}$ \ be a complex linear subspace of real dimension \ $2k$. \ Then
\begin{equation}
        vol (V\cap \phi (B_{2n} (r))) \ \leq \ c_{2k} r^{2k}, \hspace{10mm} 1 \ \leq \ k \ \leq \ n 
\end{equation}
where \ $c_{2k}$ \ is the volume of the unit radius ball in \  $\mathbb{R}^{2k}$. \ The equality holds if and only if the linear subspace \ $\phi^{-1}(V)$ \ 
is complex. The ``complex"  refers to the existence of an (almost) complex structure which pairs together the canonical positions and momenta in each 
symplectic 2-plane. Such an involution \ $J$ \ on the tangent bundle \ $T\mathcal{M}$ \ of \ $(\mathcal{M}, \omega)$ \ obeying \ $J^2 = -1$ \ provides a
relation between the symplectic structure \ $\omega$ \ and the Hermitian metric \ $\mathbf{g}$ \ of \ $\mathcal{M}$ \ via
\begin{equation}     
      \omega (X,Y) \ = \ \mathbf{g}(X, JY), \hspace{5mm} \forall \ X, Y \in T\mathcal{M}  \nonumber
\end{equation}
The result is that  sections of the unit ball under symplectic maps can have arbitrarily small $2k$-dimensional volume. 
To conclude, sections of any dimension, not only 2-dimensional ones, do not present an obstruction to embeddings under linear symplectic maps.\\  

It is understood that the set of initial conditions will be hugely deformed under a generic Hamiltonian flow. Therefore the shape of 
polydiscs and ellipsoids that may encode desirable or experimentally accessible initial conditions will dramatically change under such 
symplectic flows. One question, which was  asked by H. Hofer, is whether apart from the non-squeezing theorem, there are higher dimensional
non-trivial higher dimensional symplectic capacitie) that may further constrain such an evolution. The answer is not known in general.
However for polydiscs one has \cite{Guth} that no such non-trivial intermediate-dimensional capacities that would provide additional constraints/rigidity 
exist. Many things about the conditions for  symplectic embeddings of ellipsoids are also known \cite{Schlenk}. 
More generally, but only for linear symplectic maps, \cite{AbbMatv} prove in the same work as of that in the previous paragraph, 
that the symplectic non-squeezing theorem holds for middle-dimensional ($1\leq k \leq n$, \ in \ $\mathbb{R}^{2n}$)  linear symplectic maps, 
but does not hold for general non-linear symplectic maps. For additional results in this direction, see \cite{PN, Rigolli}. The possible physical 
implications of such results are still unclear.\\
     
 
 \subsection{The significance of the large number of degrees of freedom}    
          
The above argument may give the impression that lack of time reversibility may also be observed in Hamiltonian systems of one degree of freedom. Indeed, it appears 
that because  their phase space is 2-dimensional, it can be considered to be a 2-dimensional symplectic plane and the above argument would trivially work
for the identity embedding.  Even though time-irreversible dynamical systems of one degree of freedom can certainly be constructed, for most cases 
of fundamental physical interest the above impression would be incorrect. The irreversible behavior is observed when  there is separation between the 
microscopic and the macroscopic scales and for systems with many degrees of freedom \cite{Leb1, Leb2}. 
This had been pointed out even by L. Boltzmann. The large number of degrees of freedom seems to be  necessary, for such a behavior \cite{V-book}.
We  explain why such a large number of degrees of freedom is also necessary in our approach, in what follows.\\ 

In the symplectic non-squeezing theorem, 
the ball is not only an expression of the independence of the harmonic oscillator degrees of freedom. Through the Central Limit Theorem, and based only on the
Euclidean structure of the phase space $\mathbb{R}^{2n}$ , it can also be seen as an approximation to the Gaussian behavior of a high dimensional system,
in its thermodynamic limit. This is due to the fact that a section of a ball of fixed radius and high dimension converges to a section of the Gaussian, 
as the dimension of the ball increases to infinity, which is essentially a geometric expression of the Central Limit Theorem. This view can be traced back 
to J.C. Maxwell, E. Borel and P. L\'{e}vy and has been brought recently to prominence through the work of M. Gromov and V. Milman \cite{GM}. 
We commented about some of its implications for Statistical Physics in \cite{NK1}. \\

From the geometric view of the Central Limit Theorem \cite{GM}, a ball can be seen to represent a limiting behavior of convex sets of initial conditions in phase space.
But this statement is valid only in spaces of high dimension. Therefore, even though on dynamical grounds alone someone could use the symplectic 
non-squeezing theorem to argue about time irreversibility in systems with one, or few, degrees of freedom, the sets of initial conditions that are covered by the theorem
and its corollaries are so special and so unlikely to occur in a typical physical situation, as to render such results physically irrelevant. \\

Moreover, for systems with few  degrees of freedom, it is easier to saturate the inequality (13) in the symplectic non-squeezing theorem, 
something that would render  the above analysis, which relies on the increase of the 2-dimensional area of symplectic projections in phase space,  irrelevant. 
Intuitively at least,  keeping the area of such symplectic  projections invariant under a Hamiltonian flow seems to be far more unlikely for systems 
having a large number of weakly-correlated 
degrees of freedom, hence geometric possibilities of deformations of shapes of initial conditions,  
as far more possibilities occur in such higher dimensional spaces.   \\

For systems with many degrees of freedom, one can ask whether the symplectic non-squeezing 
theorem applies to the thermodynamic limit, namely to the case of infinite dimensional phase spaces/symplectic manifolds.  
This is hard to address in its full generality, so one may wish to start with a better understanding of this issue for infinite dimensional 
linear spaces and even more so, among them, for Hilbert spaces. The theorem was proved to be valid in the linear, Hilbert space setting by \cite{Kuk}, and further 
generalized by \cite{AbbMaj} among several works in this direction, many of which pertain to partial differential equations seen as infinite dimensional 
integrable systems, hence they may prove to be  of  some physical interest upon closer examination. \\
    

\section{Conclusions and discussion}

In this work, we argued that one can ascribe the macroscopic time irreversibility of physical systems of many degrees of freedom having reversible microscopic 
dynamics, to the symplectic non-squeezing theorem, or more generally to the existence of symplectic capacities for systems having a Hamiltonian evolution.  We 
commented on the physical aspects of the set of initial conditions employed in the application of the theorem and commented on the necessity of having 
many degrees of freedom in order for its conclusion to be typical, therefore physically relevant.\\    

 The significance of the present work is that it partially reduces time-irreversibility to being a phenomenon that can be ascribed to
essentially 2-dimensional manifolds, namely, to systems having one degree of freedom. However such phase spaces have to be 
embedded submanifolds of a larger phase space for the projections onto them to be make sense.  
The advantage of our approach is that the employed symplectic 2-dimensional planes can be at most Riemannian surfaces 
for which a lot of things are known,  a fact that has allowed the proof of numerous results in dynamical systems \cite{KH, Zehn} on such surfaces.
Thus fundamental, but practically intractable, problems of systems of many degrees of freedom, may be partially reduced  to the more manageable 
case of one, or a few degrees of freedom.\\     

Looking toward the future, we would like to explore the possibility of other physical implications of the non-squeezing theorem relevant
to Statistical Mechanics. In particular,  we wish to see whether properties of the symplectic capacities can be used to somehow differentiate 
between systems with short versus long-range interactions. At a first glance, the answer appears to be negative, but far more needs to be 
understood for such a possibility to be ruled out. In particular, the interaction between the metric and the symplectic structures of phase spaces
such as through the investigation of embeddings of the pseudo holomorphic curves in high dimensional symplectic manifolds  
and their appropriate limits may prove  to be useful in this context. This can be seen as part of  an attempt to uncover the fundamental 
dynamical features that may lead us to the use  of a power-law \cite{Ts-book, Kan} versus the conventional Boltzmann/Gibbs/Shannon entropy 
in the description of the collective behavior of Hamiltonian systems of many degrees of freedom. \\ 
  

               \vspace{0mm}

\noindent{\normalsize\bf Acknowledgement:}  \ We are indebted to the referee for his/her comments that corrected some mistakes and improved the 
exposition. We are grateful to Profs.  P. Benetatos, A. Bountis, I. Haranas, E.C. Vagenas  for their support 
without which this work  would have been impossible.  \\  





\end{document}